\documentclass[12pt,letterpaper]{article}
\usepackage{amsmath,amssymb,pgf,pgfarrows,pgfnodes,float,appendix, hyperref}
\usepackage{graphicx}
\usepackage{subfigure}
\usepackage[margin=0.9in]{geometry}

\newcommand{\be}{\begin{equation}}
\newcommand{\ee}{\end{equation}}
\newcommand{\bea}{\begin{eqnarray}}
\newcommand{\eea}{\end{eqnarray}}

\title{{\rm\footnotesize \qquad \qquad \qquad \qquad \qquad \ \qquad \qquad \qquad \ \ \ \ \ \                  RUNHETC-2015-8     
SCIPP 15/07}\vskip.5in     Holographic Space-time Models in $1 + 1$ Dimensions }
\author{Tom Banks\\
Department of Physics and SCIPP\\
University of California, Santa Cruz, CA 95064\\
{\it and}\\
Department of Physics and NHETC\\
Rutgers University, Piscataway, NJ 08854\\
E-mail: \href{mailto:banks@scipp.ucsc.edu}{banks@scipp.ucsc.edu}
\\
\\
}
\date{}
\begin{document}
\maketitle

\begin{abstract} We construct Holographic Space-time models that reproduce the dynamics of $1 + 1$ dimensional string theory.  The necessity for a dilaton field in the $1 + 1$ effective Lagrangian for classical geometry, the appearance of fermions, and even the form of the universal potential in the canonical $1$ matrix model, follow from general HST considerations. We note that 't Hooft's ansatz for the leading contribution to the black hole S-matrix, accounts for the entire S-matrix in these models in the limit that the string scale coincides with the Planck scale, up to transformations between near horizon and asymptotic coordinates. These $1 + 1$ dimensional models are describable as decoupling limits of the near horizon geometry of higher dimensional extremal black holes or black branes, and this suggests that deformations of the simplest model are equally physical.  After proposing a notion of ``relevant deformations", we describe deformations, which contain excitations corresponding to linear dilaton black  holes, some of which can be considered as UV completions of the CGHS model\cite{cghs}. We study the question of whether the AMPS\cite{amps} paradox can be formulated in those models.  It cannot, because the classical infall time to the singularity of linear dilaton black holes, is independent of the black hole mass.  This result is reproduced by our HST models. We argue that it is related to the absence of quasi-normal modes of these black hole solutions, which is itself related to the fact that the horizon has zero area. This is compatible with the resolution of the AMPS paradox proposed in \cite{fw23} , according to which the compatibility conditions of HST identify the long non-singular sojourn of observers behind the horizon, with the dynamics of equilibration on the horizon as seen by a detector which has not yet fallen through the horizon.\end{abstract}

\section{Introduction}

The basic principles of HST describe the dynamics of space-time from the point of view of individual time-like trajectories.  Nested causal diamonds along the trajectory correspond to nested tensor factors in the Hilbert space of the quantum theory.   That Hilbert space is constructed as the smallest irreducible representation of a finite dimensional super-algebra.  There is at least one cyclic vector, such that the action of the fermionic generators sweeps out the entire representation space, when applied to that vector\footnote{Examples of such super-algebras are canonical fermions, and the superalgebras generated by the commutation and anti-commutation relations of the GellMann matrices.}.  These fermionic operators are identified with the eigen-spinors of the Dirac operator of the holographic screen\footnote{The HST formalism uses the phrase {\it holographic screen} to denote a different geometrical object than the holoscreens of \cite{bousso}. Bousso describes a screen as a $d - 1$ dimensional surface with a preferred null foliation.  The HST screen is just the maximal area leaf of that foliation.} of the diamond, with some sort of cutoff on its spectrum.
This is a very general version of the UV/IR relation familiar from the AdS/CFT correspondence. Larger holoscreens correspond to higher UV cut-offs.  This postulate implements the covariant entropy/holographic principle\cite{BHtHFSB}, which Jacobson\cite{ted} showed implied that in dimension $\geq 2 + 1$, the Einstein equations were the hydrodynamic equations of the system\footnote{Jacobson's argument ``misses"  the cosmological constant (c.c.).  Fischler\cite{westschrift} and the present author \cite{tbcc} argued that the c.c. is an asymptotic boundary condition, controlling the relation between large proper times and large areas, rather than a contribution to local hydrodynamics.}.  Jacobson's argument also gives the cleanest proof that the covariant entropy bound refers to the logarithm of the dimension of the Hilbert space, because the entropy referred to is that seen by the infinite temperature Unruh trajectory.

In $1 + 1$ dimensions this whole line of thinking seems to encounter an insuperable object.  The intersection of a space-like hypersurface with the boundary of a causal diamond is two points.  There is no area.  Not coincidentally, in $1 + 1$ dimensions the Einstein equations are vacuous.  Every metric has vanishing Einstein tensor because the Einstein-Hilbert action is a topological invariant.  
The simplest way to write field equations for geometry is to introduce a scalar field $\phi$, often called the dilaton, and an action of the form
\begin{equation} \int \sqrt{- g} [- \frac{1}{2} Z(\phi ) g^{\mu\nu} \partial_{\mu} \phi \partial_{\nu} \phi  + F(\phi ) R + V(\phi )] . \end{equation} One of the three functions, $Z,F,V$ can be eliminated by a field redefinition, leaving two free functions of the dilaton. 

HST gives us a physical interpretation of the dilaton field.  From the HST point of view, we need, in the absence of holoscreen area, a separate function to describe the entropy of causal diamonds.  Indeed, many of these Lagrangians have black hole solutions, and the black hole entropy is always a simple function of the value of the dilaton at the horizon.  Another general feature is that gauge fixing eliminates all propagating fields in these models, and the entire phase space is parametrized by a finite number of parameters.  For the Lagrangians with linear dilaton vacuum solutions, which will be our concern in this paper, the only parameter is the black hole mass.  These linear dilaton black holes thus have no quasi-normal mode spectrum, a fact which will be of importance in the sequel.

There is another illuminating way to view the fact that an extra scalar is needed to describe the entropy, since many of the Lagrangians can be derived by starting from a higher dimensional space-time and throwing away all but one mode of the metric in higher dimensions, for example the radius of spheres in a higher dimensional Minkowski space, in which $r$ is a radial coordinate.  Indeed, as we will recall in Section 2 , dilaton gravity can be derived as the decoupled near horizon limit of certain extremal black holes in string theory.  We will interpret the (HST completion of the) CGHS\cite{cghs} model, based on this correspondence, as the first example of a duality between a gravitational theory and a holographic non-gravitational theory. We will argue that the full quantum theory of these low energy effective field theories is given by ``relevant" modifications of the fermionic field theories dual to $\hat{c} = 1$ RNS strings\footnote{Indeed, we considered entitling this paper ``A New Hat for the CGHS Models". We will not attempt, in this paper, to make a detailed catalog of matches between specific extremal dilaton black holes, and specific matrix models but instead try to define what {\it relevant deformation} means in the context of the existing matrix models. }.   The vast landscape of string compactifications, which support linear dilaton black holes leads us to expect that there will be many consistent $1 + 1$ dimensional models of quantum gravity, with varying massless field content.  We will try to capture a subset of these models as simple deformations of the $\hat{c} = 1$ model, and extension of its field content.

In HST the physical DOF are fermions $\psi_k$\footnote{I will leave the exploration of more complicated superalgebras\cite{hstsummary} to another paper by another author.} .   The entropy of causal diamonds is $K {\rm ln 2} $, where $K$ is the total number of fermionic oscillators, and in space-time dimension $d \geq 3$ that number represents, via the UV/IR connection, the area of a holoscreen.  Systems with a finite number of fermionic oscillators do not have a scattering interpretation, so the system can only describe something with the causal structure of $1 + 1$ dimensional Minkowski space\footnote{In a $1 + 1$ dimensional Lagrangian model of gravity and scalar fields, we often make a field redefinition so that a metric with the causal structure of Minkowski space, {\it is} just Minkowski space.}, and a scattering theory, if the entropy of some causal diamonds can go to infinity. 
Consider some time-like trajectory in $ 1 + 1$ Minkowski space. As time goes on, the trajectory's diamond has more entropy.  In $1 + 1$ dimensional language, this means that the dilaton field must be varying in space-time. It's important to emphasize that this must be a property of the space-time that represents {\it the ground state } of the quantum system.  When we use the word ground state, we mean that the system has a time-independent Hamiltonian, which means that the space-time has a global time-like Killing vector\footnote{Strictly speaking, we only need an asymptotic Killing vector, but the black hole solutions, which have a Killing horizon, are supposed to represent the hydrodynamics of high entropy excitations of the ground state.  The ground state has a global time-like Killing vector.}   Indeed, many of the $1 + 1$ dimensional Lagrangians above have solutions where the metric is Minkowski, and the dilaton depends only on space, and goes to infinity at infinity. Such space-time backgrounds, in which the entropy goes to infinity at $ r = \infty$, might be scattering theories because the asymptotic Hilbert space can have a Hamiltonian with continuous spectrum.

The entropy as a function of $r$ can have only a single minimum.  This follows from the a consistency relation between causality and conformal factor in the HST description of space-time.  A smaller causal diamond, must, by causality, contain fewer DOF than a large one, because the existence of regions in the larger diamond, which are space-like separated from the smaller one, shows that the operator algebra of the smaller diamond is a proper sub-factor of the larger algebra.   The single minimum thus represents the spatial point of minimal entropy.

There seems to be a choice of whether the one dimensional space is an infinite line or a half line, but this is illusory. In the higher dimensional picture that applies to linear dilaton black holes, it's natural to think of it as a half line.
From the $1 + 1$ dimensional point of view, the difference between the infinite and half infinite line is a matter of labeling.  In one dimension, on an infinite line with inhomogenieties, we have left and right moving asymptotic scattering states at both $\pm$ spatial infinity.  The definition of left and right is a convention, and we can always map a two sided problem into a one sided problem with more asymptotic labels.
In the HST context, the fold is naturally made at the point of minimum entropy.
There is no need for a reflection symmetry about that fold, but the models we study do have such a symmetry.

Now consider the causal diamonds along a Minkowski geodesic at rest in the frame defined by the entropy function $S(r)$, and situated at the point of minimal entropy, $r = 0$.  We will call this the {\it minimal entropy trajectory}, though in the limit of infinite proper time its causal diamonds see the full Hilbert space of the system.  In that limit, the Hamiltonian becomes time independent. The $r$ coordinate of the holographic screen at time $t = r$, is $r$.  In linear dilaton systems, describing the near horizon region of a large class of extremal black holes, the entropy function asymptotes to $S(r) \rightarrow e^{\frac{r}{\ell}} , $ where $\ell$ is a length scale defined by the cosmological constant in the graviton-dilaton effective action.  We will identify this length scale as the Planck length $\ell = L_P$.  It is the only parameter with dimensions in the $1 + 1$ dimensional gravitational action.

Extremal dilaton black holes have a global time-like Killing vector, but the smallest perturbation gives them a horizon at a finite value of $r$.  For any slight deviation from extremality, the time-like Killing vector at infinity, acts like a Lorentz boost in the near horizon region.  
We now come to a major assumption, which is justified by its correspondence with the well established matrix models of $\hat{c} = 1$ RNS strings. We assume that the classical statement that the Hamiltonian $H$ in the asymptotic Hilbert space acts like 
a boost, is encoded in the quantum theory through the existence of two {\it near horizon null plane coordinate operators, $u$ and $v$} such that 
\begin{equation} [h,u] = u , \ \ \ \ \ [h,v] = - v  . \end{equation} These operators are related, for large (positive for $u$, negative for $v$) values of their spectra to classical light front coordinates $t \pm r$ on ${\cal I}_{\pm}$ , which is the infinite entropy region where scattering states are defined.  We'll see precisely what the asymptotic map is below. 

The relation between the operators $h$ and $H$ will be given by second quantization. In higher dimensional HST, physical DOF are associated with sections of the spinor bundle on the holographic screen, with a Dirac cutoff related to the area of the screen\cite{tbjk}.  This encodes the higher dimensional area/entropy connection. In $1 +1$ dimensions, where the screen is a point, we instead employ fermions, which depend on the variables that determine where the entropy is concentrated.
The fermionic DOF can thus be thought of as functions $\psi_i (u) $, {\it or} $\psi_i (v)$ of the light front operators, which approach ordinary asymptotic quantum fermion fields in the appropriate limit.  \footnote{In passing we note that even without the input from HST, we would have chosen fermion fields to represent asymptotic massless particle states in $1+1$ dimensions.  All such particles are either spinless bosons, or fermions.  Scalar and complex fermion fields are equivalent, because of bosonization, and the fermion representation avoids the need to deal with solitons. It is also more general, since Majorana fermions cannot be bosonized into free bosons}.

The aim of a model of quantum gravity in linear dilaton backgrounds is to construct a unitary S-matrix on the asymptotic Hilbert space. We've emphasized that the fermions can be thought of as functions of {\it either} $u$, or $v$, but not both, since the spectra of these operators parametrize the asymptotic past and future boundaries of space-time.  Thus, the two different representations of the asymptotic variables should be related by the S-matrix, and not independent of each other.   The simplest possible model of this type will have an S-matrix which just transforms $u$ into $v$, and a single fermion field $\psi (v) = \psi (s^{\dagger} u s) , $ where $s$ is a ``one body S-matrix".  We'll explain the phrase in quotes in a moment, but {\it cognoscenti} of old matrix model lore will recognize exactly what we're talking about.  $v$ will commute with $u$ only if $s$ does, and this corresponds to a trivial $S$ matrix, $S = 1$.  

The simplest non-vanishing commutator between $v$ and $u$, consistent with their commutators with $H$, is 

\begin{equation} [u, v] = - i , \end{equation}  which also specifies the normalization of these operators.  The sign in this commutator is a convention, which is chosen to agree with \cite{maldaseibflux}. We've renamed the operator they call $s$ as $v$, to coincide with the standard convention for light front coordinates. This is the $1 + 1$ dimensional version of the commutator between light front coordinates postulated by 't Hooft\cite{tH} , to account for the scattering between in-falling particles and outgoing Hawking radiation.  Note that although 't Hooft worked in the context of Schwarzschild black holes in $4$ dimensions, his arguments apply to near extremal linear dilaton black holes, for {\it any} deviation from extremality, and should thus be imagined valid in the extremal limit as well.  We emphasize that the extremal higher dimensional black hole is the ground state of our $1 + 1$ dimensional model of the decoupled near-horizon limit.

We are now done.  The unique (up to a unitary conjugation) irreducible unitary representation of the algebra of $h, u$ and $v$ (more properly of the unitary group obtained by exponentiating these operators) is that in which $u$ and $v$ are Fourier conjugate multiplication/derivative operators on ${\cal L}^2 (R)$ and 
\begin{equation} h = \frac{1}{2} (uv + vu) . \end{equation}
The minimal fermionic Fock space representation of the algebra has 
\begin{equation} U = \int du \psi^{\dagger} (u) u \psi (u) , \end{equation}
\begin{equation} V = i \int du \psi^{\dagger} (u) \frac{\partial}{\partial u} \psi (u) , \end{equation}
\begin{equation} H  = \frac{1}{2} \int du \psi^{\dagger} (u) (uv + vu) \psi (u) . \end{equation}
Introducing $u = (p + \lambda)$, $v = p - \lambda$, where $p$ and $\lambda$ are canonical conjugates, we see that we've derived the fermionic field theory of the $0B$ RNS string in a linear dilaton background.  We'll elaborate on this in Section 2, and explore in particular the localization of entropy in that picture. This investigation will derive the linear dilaton background from the matrix model.  The connection between the $c = 1$ model and the Dray-'t Hooft shock wave calculation was noted long ago in \cite{verlindefriess}, but the full calculation of the S-matrix from this point of view was done in \cite{kk}. The latter authors did not notice the connection of their work to the gravitational shock wave calculation.

In Section 3. we'll discuss the concept of {\it relevant perturbation} of the 0B matrix model and argue that a class of perturbations of the form
$$ \int d\lambda d\lambda^{\prime} \psi^{\dagger} (\lambda ) M(\lambda , \lambda^{\prime} ) \psi (\lambda^{\prime} ) , $$ with a kernel suitably smooth, and localized near the maximum of the one body potential, deserve that appellation. We'll also generalize the model to a large number of fermions, making contact with the CGHS\cite{cghs} model, and argue that this is the analog of taking the AdS radius much larger than the Planck or string scales\footnote{The simplest relevant perturbation is the familiar tachyon condensate of the $0B$ model, which, when taken large, gives us a weak string coupling expansion.  In this paper, we will mostly discuss the opposite limit, where the tachyon condensate is absent or small, and the Planck and string scales coincide approximately.}  .  We'll argue that in this limit, a large class of perturbed models have meta-stable excitations, whose behavior resembles that of black holes.  

In Section 4. we'll examine the AMPS\cite{amps} argument in the large $N$ models of Section 3.  We'll see that some of the AMPS assumptions are incorrect because of the existence of a large number of closely spaced levels, which cannot be described in the low energy effective field theory (which has, as usual, a unique low energy vacuum state).  These levels are localized near the point where the original matrix model has low entropy.  The relevant perturbation does not by itself change this much, but for large $N$, we can fill the few metastable levels with a large number of different fermions.  To create such a state, starting from the vacuum, one must send in a large number of incoming fermions.  We'll argue that this is analogous to the fact that a higher dimensional black hole has to have energy much larger than the Planck scale.  In the $N \rightarrow\infty $ limit, we can describe the incoming fermion state in terms of a classical density wave, recovering the classical solution of the CGHS model, which creates a large black hole in what appears to be a reliable semi-classical approximation.   In HST language this is equivalent to saying that the entropy computed from the matrix model is large at that point.  The relationship between matrix model and space-time coordinates shows that this entropy is attributable to meta-stable fermion states localized near the origin of the matrix model coordinate.  These states modify the relationship between the classical space coordinate, and the eigenvalue coordinate of the matrix/HST model.   

It turns out however that we cannot really address the AMPS paradox in these models.  The in-fall time from the horizon to the singularity of a linear dilaton black hole does not scale with the black hole entropy/mass.  Like the temperature, it is of order the Planck scale.  If we follow the interpretation of the interior of the black hole advocated in \cite{fw23}, according to which the interior encountered by a particle system falling past the horizon, is the interpretation along a time-like trajectory close to that system, of the equilibration of that system on the horizon seen along trajectories which fall into the black hole at later times.  According to that interpretation, the long period (in Planck units) before that system is disturbed by the singularity is the mirror of the long time that it takes the horizon to begin to come into equilibrium.

The black holes of discussed above are analogous to domain walls in AdS/
CFT .  They do not represent states in the quantum theory of the undeformed model. We also examine an alternative class of
perturbations, analogous to double trace perturbations in large $N$ models.  These models have fermion interactions, but no change to the bare one body potential.  We argue that the deformed potentials of the simpler models arise as metastable soliton solutions of the Hartree equations.  These models come closest to what we expect from higher dimensional models of black hole formation and evaporation.  They differ from a theory of multiple decoupled copies of the 0B matrix model only by interactions between different fermions, localized in the low entropy region of space-time.  If those interactions are attractive, high entropy meta-stable states of the theory form. They are localized near the erstwhile low entropy region, and produce a horizon in the low energy effective field theory.   These states slowly evaporate, preserving unitarity, and approximate locality.  

After this work was completed, I became aware of the paper \cite{dubflaug} which also studies a $1 + 1$ dimensional theory of quantum gravity.  The two approaches appear to be radically different, and certainly the formulae for the S matrix do not coincide.  Nonetheless, some of the remarks in \cite{dubflaug} about black hole behavior and the possibilities of thermalization of entangled initial states in integrable systems are quite similar to ideas presented here.

\section{Summary of Earlier Work on Linear Dilaton Models}

\subsection{Low Energy Effective Field Theory}
The equations of motion of the gravitational Lagrangian 

\begin{equation} L = \frac{1}{2\pi} \int d^2 x\ \sqrt{-g} e^{ - 2\phi} [R + 4 (\nabla\phi)^2 + \frac{4}{L_P^2} ] , \end{equation} have conformal gauge solutions (we use metric conventions in which time-like vectors have positive inner product)
\begin{equation}  g_{uv} = - \frac{L_P^2}{2} e^{2\rho} = L_P^2 (M L_P - uv)^{-1} = L_P^2 e^{2\phi} . \end{equation}
 For $M > 0$ this is a black hole, with horizon at $uv = 0$.  The $M = 0$ solution is called the {\it linear dilaton vacuum} .  The physical region of coordinates for this solution is $u \geq 0$ and $v \leq 0$.  By going to coordinates  proportional to
 ${\rm ln} u$ and $ {\rm ln} ( - v)$, we see that the metric is that of Minkowski space, while the dilaton field is a linear function only of the space-like coordinate 
$r = L_P [ {\rm ln} u + {\rm ln} ( - v) ]$.  The boost isometry, which rescales $u$ and $v$ in opposite directions, is translation of the time coordinate $t = L_P [ {\rm ln} u - {\rm ln} ( - v) ]$.  Overall scaling of both $u$ and $v$, the spatial translation in Minkowski space, leaves the line element
$ ds^2 = 2 g_{uv} du dv $, invariant for $M = 0$.  For non-zero $M$, the same coordinate transformation rescales $ML_P$, so that space-time intervals are independent of this parameter.  

The physical region of the coordinates when $M \neq 0$ extends to the singularity at $uv = ML_P$ and the proper time transit from the horizon ($uv = 0$) to the singularity is $o(1)$ in Planck units.  The authors of \cite{rabinovicietalwitten} and \cite{cghs} showed that $M$ is properly interpreted as the black hole mass\footnote{The linear dilation black hole solution was first constructed as a string background in Euclidean space, by Elitzur, Forge and Rabinovici. Witten constructed it using the coset methods of Bardacki {\it et. al.} and proposed the black hole interpretation. CGHS found the solution independently by studying the near horizon region of higher dimensional dilaton black holes.} .  For our purposes 
it is more important to understand the relation of $M$ to the black hole entropy.
This can be gleaned from the dimensional reduction described in\cite{horstromgidd}.  One finds, not surprisingly, that this is give in terms of the dilaton field, which parametrizes the area of a co-dimension 2 surface in the higher dimensional space-time.
\begin{equation} S = e^{ - 2\phi (uv = 0)} = ML_P .  \end{equation}   This equation implies that the temperature of the black hole is mass independent and of order $L_P^{-1}$.

We can create a black hole from the linear dilaton background by sending in an incoming shock wave, with stress tensor

\begin{equation} T_{uu} = \frac{a}{L_P^2}  \delta (u - u_0)  . \end{equation}
The resulting metric and dilaton fields are

\begin{equation} e^{- 2\rho} = e^{-2\phi} =  - a (u - u_0) \theta (u - u_0) - uv . \end{equation}
Shifting $v\rightarrow v - a $, we see that we've produced a black hole with
$M L_P = a$.  The shift of $v$ is the analog of that predicted by 't Hooft and Dray.  Following 't Hooft's argument we've interpreted it as a non-vanishing commutator between $u$ and $v$ in the quantum theory.  Finally, this calculation exhibits the logic that led us to identify the length scale associated with the two dimensional c.c. as the Planck scale.  We get a semi-classical black hole only when its mass is large in Planck units.  

The CGHS model coupled this gravitational system to free massless quantum fields.   For a small number of fields there is no real justification for the black hole formation calculation we have just done.  In the language of effective field theory, the states of the matter fields which could generate a classical shock wave with $M \gg M_P$ would be states in which higher dimension operators scaled by the Planck length became important.  For models with a large number of fields, we can construct high energy pulses without going beyond the rules of effective field theory.
As we'll see below, this argument is not exactly correct, because there is a large class of fully quantum linear dilaton models, for which the only corrections to the effective field theory occur in the region of strong coupling.
Nonetheless, the conclusion of the argument, that we need a large number of matter fields to form a semi-classical black hole, is indeed correct.

\subsection{Linear Dilaton Vacua in String Theory}

Linear dilaton solutions of low energy Lagrangians for quantum gravity have appeared in at least two different contexts in the string theory literature.  Strictly two dimensional approaches include the bosonic and Type 0A,B perturbative strings\cite{bose0AB}, and Witten's two dimensional black hole\cite{witten} .  In contrast, the CGHS model\cite{cghs} derived these two dimensional models as conjectured decoupling limits of the near horizon dynamics\footnote{The language used here is somewhat anachronistic, but it seems clear that \cite{cghs} was an unwitting precursor of the AdS/CFT correspondence.}  of certain extremal black holes discovered in higher dimensional super-string models with $4$ or more asymptotically flat dimensions.  We would conjecture that in fact all such models arise from the latter context.  A full catalog of linear dilaton throats of higher dimensional models does not exist, and we will not attempt to construct one in this paper.

The perturbative string theory approach to these models makes sense only in a certain limit.  The linear dilaton solution implies that the string coupling blows up in the interesting region where black hole formation might be occurring.  A perturbative string regime can be enforced by introducing a perturbation which prevents asymptotic states from approaching the strong coupling region of space-time.  This is the tachyon condensate, which, from the matrix model/fermion point of view is an adjustment of the fermi surface. This prevents fermions from approaching the maximum of their one body potential.  While the perturbative string perspective was crucial to establishing the legitimacy of these models, it introduces a clumsy machinery, which obscures much of the important physics.  Even the computation of perturbative scattering amplitudes is much harder in string theory language than in the direct expansion of the fermionic answers\cite{polchinskiwhatis}\footnote{One unresolved issue is the derivation of the (in)famous {\it leg pole} phases from a purely fermionic perspective.  It's clear however that much of the physics of the leg poles is connected to special features of the weakly coupled regime and the fact that there is a string length scale, unconnected to entropy counting, in this regime.}.   More importantly, string perturbation theory obscures the instability of the bosonic string, and, as we will see, the nature of $1 + 1$ dimensional black hole physics.  To make a not very precise analogy: in \cite{tbwfbh} we argued that weakly coupled string theory in flat space had at least three pre-asymptotic regimes, which interfere with a direct confrontation with black hole evaporation.  These are the perturbative string regime, the Gross-Mende regime\cite{grossmende} and the regime of D brane and D instanton physics.  For this reason, we will turn off the tachyon condensate for the bulk of this paper.

The stringy approach to these $1 + 1$ dimensional models derives from 't Hooft's\cite{tHlargeN} connection between the genus expansion of large $N$ matrix models and the genus expansion of string perturbation theory.  At finite orders in $1/N$ the Feynman diagram expansion in the 't Hooft coupling $g_t$ has a finite radius of convergence.  As we approach the singular point in 't Hooft coupling, two things happen.  At each order in the genus expansion, large diagrams, which one might imagine to be well approximated by continuum random surfaces, become important.  Simultaneously (and this requires some analysis which we will not reproduce here), the suppression of higher genus diagrams by powers of $1/N$ is compensated by their more singular behavior as $g_t \rightarrow g_t^*$ .  One searches for, and finds, universal behavior in the {\it double scaling limit} in which $g_t \rightarrow g_t^*$ as $N$ goes to infinity, in such a way that the suppression of higher genus diagrams remains finite, parametrized by a continuum string coupling $g_s^{2g - 2}$.  

The simplest example of this type, with a quantum mechanical interpretation, is that of a single $h \times h$ Hermitian matrix $H$ with (Euclidean) Lagrangian
\begin{equation} L = \frac{\beta}{2 } {\rm Tr}\ [\dot{H}^2 + V(H) ] . \end{equation} The $U(h)$ symmetry of this model should be interpreted as a gauge symmetry: only singlet states are physical.  This is best understood by realizing the matrix model as the dynamics of unstable $D0$ branes of the type 0B string theory\cite{bose0AB} .  The fact that this model completely captures the dynamics of the closed string sector of the theory shows that the matrix model is an avatar of the AdS/CFT correspondence.  The fermionic nature of the $D0$ branes comes from the peculiarities of matrix dynamics. The matrix $H$ can be expressed in terms of the unitary which diagonalizes it and of its eigenvalues.  The latter are the gauge singlets, and projecting onto the singlet sector, introduces a Van der Monde determinant into their functional integration measure.  The square root of the determinant appears in the quantum wave functions for these variables, and the ambiguous $\pm$ sign in the square root gives rise to the $( - 1)^F$ gauge symmetry that characterizes fermions.  The remaining $S_h$ statistical gauge symmetry is the residual action of $U(h)$ on the eigenvalues. This agrees with the analysis of the statistics of $0B$ $ D0$ branes in continuum string theory.

The singlet sector of the matrix model is thus the quantum mechanics of $M$ non-interacting fermions, with a $1$ body potential $V(\lambda )$.  The critical behavior of the large $h$ limit occurs when the chemical potential approaches a maximum of the potential, which is generically quadratic. The more finely tuned multi-critical points with higher order maxima have not been explored in great detail.  The double scaling limit is one in which $h \rightarrow \infty$ and the potential is just an inverted harmonic oscillator. We obtain a fermionic field theory, with a complex fermion field $\psi (\lambda )$. The Lagrangian of the double scaled theory is
\begin{equation} L = \int_{-\infty}^{\infty} d\lambda\ \psi^{\dagger} (\lambda, t) [i\partial_t - \frac{\omega}{2} (p^2 - \lambda^2 )  - \mu ] \psi (\lambda , t) . \end{equation} Here $p = \frac{\partial}{i\partial\lambda} $.  The bosonic string corresponds to the {\it obviously} unstable situation of only one side of the Fermi sea filled.   This is obscured by taking the limit $\mu \gg \omega$, in which the scattering matrix has a perturbative expansion around $S = 1$\footnote{The bulk scattering matrix is exactly equal to $1$ in this model with $\frac{\mu}{\omega} \gg 1$\cite{grossklebanov}.  Scattering occurs only near the tachyon wall, where the string coupling becomes large.} The stable situation of equal chemical potentials on both sides of the maximum corresponds to the $0B$ NSR string, while a finite imbalance is a flux vacuum of $0B$\cite{maldaseibflux}.

The most transparent description of scattering in this model was invented by Alexandrov Kazakov and Kostov\cite{kk} and is lucidly described and extended in the appendix to\cite{maldaseibflux} .  We begin by introducing the Hermitian one body operators
\begin{equation} v \equiv \frac{p + \lambda}{\sqrt{2}} , \end{equation}
\begin{equation} u \equiv \frac{p - \lambda}{\sqrt{2}} , \end{equation} which satisfy
\begin{equation} [ v, u ] = i . \end{equation}
The Hamiltonian has the form
\begin{equation} H = \int du \psi_u^{\dagger} (u) [\omega (uv + vu) - \mu] \psi_u (u) , \end{equation}
\begin{equation} H = \int dv \psi_v^{\dagger} (v) [\omega (uv + vu) - \mu] \psi_v (v) . \end{equation}
We've used the subscripts on the two field operators to emphasize that they are different functions of their arguments, related in the usual way by Fourier transform.
In the two different representations, one of the operators $u,v$ multiplies the field by the appropriate variable, while the other is proportional to the derivative operator in that variable.  The one body Hamiltonian is, up to a constant shift, proportional to the scaling derivative $v\partial_v$ or $u\partial_u$ .  Equivalently, these are the derivatives w.r.t. to the logarithms of the variables $u$ or $v$, and the constant shifts are precisely what's required to turn the Hamilton into that of a relativistic chiral fermion with a chemical potential.  The anti-symmetry of the commutation relation under interchange of $u$ and $v$ translates in the language of relativistic fermions into the statement that the $u$ fermions are left movers and the $v$ fermions right movers\footnote{At the end of the day, our picture is that of an extremal black hole metric on a half infinite line.  We choose the convention that incoming particles on this half line move to the left, and outgoing particles to the right.
The coordinates in which the Lagrangian looks exactly like that of free Weyl fermions are the two dimensional analogs of tortoise coordinates for the black hole.}

The energy eigenstates of the one body Hamiltonian, with eigenvalue $ - \epsilon  \omega $ are pure power laws and are singular at $u$ or $v  = 0$.  This leads to a doubling of the spectrum, which is of course obvious in the $\lambda$ representation where the Hamiltonian is a second order differential operator.  Another way of seeing this is to note that the transition to ln $u$ or ln $v$ must be done separately for positive and negative values of the variable.  Thus, although the free relativistic fermions appear to live on Minkowski space, one end of that space maps to infinite values of $u$ or $v$ and the other to zero.  The two signs of $u,v$ correspond to two species of complex $1 + 1$ dimensional Weyl fermions.

To understand the meaning of the two different representations, we note that $u$ and $v$ scale in the opposite direction under the action of the Hamiltonian.  Thus ln $|u|$ and ln $|v| $ are {\it shifted} in the opposite direction.  As a consequence, the $u$ modes represent fermions that are traveling ${\it in}$ towards the maximum of the potential, while the $v$ modes are traveling outward.
More detailed arguments for this identification can be found in \cite{maldaseibflux} .  {\it We conclude that the scattering operator maps the $u$ representation into the $v$ representation of the fields.}

The precise formulae defining the relativistic fermion fields and the S-matrix are

\begin{equation}  \Psi^{(in)}_{\pm} (r,t) = e{\frac{r\omega}{2}}\psi_u ( u = \pm e^{r\omega} , t) ,\end{equation}
\begin{equation}  \Psi^{(out)}_{\pm} (r,t) = e{\frac{r\omega}{2}}\psi_v ( v = \pm e^{r\omega} , t) .\end{equation}
The Lagrangian is
\begin{equation} {\cal L} = \int_{-\infty}^{\infty} dr\ \Psi^{(in)\dagger}_{i} (r,t) (i\partial_t - i\partial_r + \mu) \Psi^{(in)}_{i} (r,t) . \end{equation}
\begin{equation} {\cal L} = \int_{-\infty}^{\infty} dr\ \Psi^{(out)\dagger}_{i} (r,t) (i\partial_t + i\partial_r + \mu) \Psi^{(in)}_{i} (r,t) . \end{equation}  The label $i$ runs over the values $\pm$ .  Although these represent modes defined for different signs of $u$ or $v$, they end up being flavor labels in the space-time representation.  

The S-matrix is implicit in the transformation between in and out fields.  This is given by
\begin{equation} \Psi^{(out)}_a (r,t) = \frac{\omega}{\sqrt{2\pi}} \int ds\ [e^{\frac{\omega(r + s)}{2}} \Sigma (\omega (r + s) )_{ab} \Psi^{(in)}_b (s,t) . \end{equation}  The matrix $\Sigma (y) $ is given by
\begin{equation} \Sigma (y) = {\rm exp} [e^{i y}] + {\rm exp} [e^{- i y}] \sigma_1 . \end{equation}
The peculiar double exponential is just the translation of the Fourier transform relation between $u$ and $v$, into the logarithmic coordinates $r$ and $s$.

The key physics of the model is given by the quantum mechanically conjugate relation between the incoming and outgoing null coordinates $u,v$.  Many years ago, 't Hooft and Dray\cite{tHDray} calculated the scattering of two Aichelburg-Sexl waves and showed that one important effect was a mutual shift in the incoming/outgoing coordinates.  Later, 't Hooft\cite{tH} interpreted this as a canonical conjugate relation between null coordinates, so that the transition matrix was the Fourier transform.  These calculations were done in four dimensions and the dynamics was complicated by dependence on the impact parameter of the collision.  

A two dimensional version of the 't Hooft-Dray calculation appeared in the seminal paper\cite{cghs} (CGHS), and we have reproduced it above.  These authors showed that an incoming shock in the linear dilaton vacuum produces a black hole, {\it but also a shift in the conjugate null coordinate} (a shock in $x^+$ produces a shift in $x^-$ and vice versa).  We claim that the matrix model gives an exact quantum mechanical justification for 't Hooft's black hole S-matrix ansatz, and that very little else is going on in that model.  Indeed, the CGHS shock wave calculation is done in conformal gauge.   The conformal gauge is related to the linear dilaton gauge, where the dilaton depends only on the space coordinate, by an exponential transformation, exactly analogous to the change of variables relating $u$ and $v$ to $r$.  The only difference is that
in the matrix model, as in 't Hooft's ansatz, the conformal gauge null coordinates are non-commuting quantum operators, obeying the Heisenberg algebra.  

Note also that 't Hooft postulated a full S matrix of the form
\begin{equation} S = S_{out} S_{hor} S_{in} , \end{equation} where only the central factor was given by the simple Fourier transform relation between incoming and outgoing null coordinates.  This was motivated by the necessity for accounting for other scattering, which has nothing to do with the black hole, but it also involves the translation between the non-commutative near horizon null coordinates, and ordinary commuting coordinates near the conformal boundary of space-time.
In the $\hat{c} = 1$ model, only the latter is necessary.

Of course, it is well known that the  $\hat{c} = 1$ model does not really have  black holes .  This, as we shall see, is related to the fact that there is very little entropy localized near the maximum of the potential.  However, the 't Hooft commutator algebra is not really related to the formation of a black hole.
In the CGHS solution the shift of the null coordinate and the additive constant in the conformal factor are independent features of the solution for the metric in the presence of an in-falling massless wave.   In order to construct a quantum model with states corresponding to the black hole solutions of linear dilaton gravity, we will have to modify the $\hat{c} = 1$ model in two ways.
In the next section we will introduce the notion of a {\it relevant perturbation} of the $\hat{c} = 1$ model and describe a large class of models, which have scattering matrices similar to that of the 0B matrix model.  To obtain semi-classical black holes, we will also have to modify the model by introducing a large number $N$ of fermions.  This is quite closely connected to the large N models of CGHS, and may be considered as a quantum completion of those models.  We'll argue that taking $N$ large plays the same role here as is does in the AdS/CFT correspondence:  it's only in this limit that semi-classical notions of locality and effective quantum field theory near the black hole horizon make (approximate) sense.

We end this review of the $\hat{c}= 1$ theory by pointing out an important symmetry of the problem.  Maldacena and Seiberg\cite{maldaseibflux} carefully studied the unitary symmetries of the model, but neglected to mention the anti-unitary time reversal transformation. This is implemented by doing the (symplectic) duality transformation exchanging $x$ and $p$ (or $u$ and $v$) without taking Hermitian conjugates of the field.   This takes the Hamiltonian $H$ to $- H$, which reverses the flow of time.  As usual, it is implemented by an anti-unitary operator, because the spectrum of the second quantized Hamiltonian is bounded from below.  The S-duality transformation of Maldacena and Seiberg is CP in space-time.   The shift of the fermi energy away from the top of the potential, which, if it is large compared to $\omega$, can generate a weak coupling string theory expansion, breaks CP but preserves T and so violates CPT.  This is manifest in the scattering states, where it acts as a chemical potential for fermion number.

\section{Relevant Deformations of the $\hat{c} = 1$ Model}

\subsection{One Body/Single Trace Deformations}

In \cite{groundring} Witten exhibited a class of discrete BRST invariant operators of the bosonic string in the linear dilaton background, and showed that their algebra was the algebra of functions on the matrix model phase space.  A similar analysis can be done for $\hat{c} = 1$. According to Seiberg's\cite{natiliouville} classification of macroscopic and microscopic operators in Liouville theory, these ``discrete states" should be considered 
perturbations of the background, rather than states  in the quantum theory.

Recall that in asymptotically flat, Poincare invariant string models, all perturbations correspond to scattering states of the theory.  In anti-deSitter space we make the distinction between normalizable and non-normalizable states, with the latter corresponding to operators whose correlation functions constitute the boundary data of the theory.  The Seiberg classification of Liouville world sheet operators is a precise analog of this.  In AdS/CFT, we make a further distinction between relevant, irrelevant and exactly marginal operators.  The irrelevant operators only make sense as infinitesimal perturbations.  Their point separated correlators are well defined, but if we try to treat the integrated operators as defining finite perturbations of the theory, then there are an infinite number of contact terms that need to be supplied.  Generically, the OPE tells us that any such perturbation in fact requires all such perturbations to be turned on.  This can be done at any finite order in perturbation theory, but the result is not summable.

The situation is different for relevant and exactly marginal perturbations, where the finite perturbations correspond to new space-time backgrounds, which either change the parameters characterizing the AdS space of the original model, or introduce domain walls in the Poincare patch.  Our aim in this section is to generalize the notion of relevant operators to the $\hat{c}= 1$ matrix models.  The key feature of such perturbations in AdS/CFT is that they do not change the asymptotic structure of space-time.

The matrix model representatives of Witten's ground ring operators are
\begin{equation} \int du \psi_u^{\dagger\ i} (u) u^m v^n \psi_u^i (u) , \end{equation} where 
$ v = \frac{d}{i du} $.  Of course, there's an equivalent presentation in terms of the field $\psi_v $.  The asymptotic regions are $u \rightarrow \infty$ and $ v \rightarrow \infty$, and it's clear that all of these operators blow up there.  Their correlators require regularization and they cannot be integrated to finite perturbations.  Remarkably though, in this $1 + 1$ dimensional model, there are combinations of the ground ring operators which do have the properties of relevant operators.  Indeed, the monomials are formally a basis for the space of all functions of $u$ and $v$.  It's easy to construct functions which are confined to the region $u \sim v \sim 0$.

A simple class of examples is $r(\beta , k) = e^{ - \beta (u^2 + v^2)^k}\footnote{In passing we note that the operators $uv + vu$ and $u^2 \pm v^2$ satisfy an $SL(2,R)$ algebra.  We suspect that this may be connected to conjectures about the description of generic horizons by chiral conformal field theories\cite{carlipetal}.} .$ For $k = 1$ this is the density operator of the right side up harmonic oscillator at inverse temperature $\beta$.  Its matrix elements between both $u$ and $v$ eigenstates fall off like a Gaussian when either of the eigenvalues goes to infinity.  We can also consider smooth deformations of the upside down oscillator potential, localized near the origin of $\lambda$ to be relevant deformations.

The intuitive physics of local deformations of the potential is easier to describe, because it is more familiar.  The operators $r(\beta , k)$ are integral kernels in the $\lambda$ coordinates, and we are quite certain they behave in a similar manner.   Our example will be a negative correction to the potential, localized near $\lambda = 0$
\begin{equation} v(\lambda ) = - \frac{\omega}{2} (\lambda^2 + v_0 (\lambda ) ) . \end{equation}
For appropriate choices of $v_0$ we will have meta-stable one body states, localized near $\lambda = 0$. By taking extreme limits of the potential, we could have many such states and make them arbitrarily meta-stable, but we will not do this.  All we require is a few such states with splitting larger than their width, and real part of the Breit-Wigner poles that are not rational multiples of each other (this is of course the generic situation).  

We now make a further modification of the system by increasing the number of complex Weyl fermions from $2$ to $N \gg 1$.  In doing this, we are following the lead of CGHS, but our analysis is somewhat different from theirs.  CGHS argued that even for $ N = 1$, an incoming wave with large fermion charge density (CGHS work with what, from our point of view, is the bosonized fermion field), created a black hole of large entropy, for which the classical equations of dilaton gravity were valid near the horizon.  The exact solution of the $\hat{c} = 1$ model shows that this is not true.  The only region of large entropy in the model is for large $u$ or $v$, the asymptotic region. Polchinski's analysis of the scattering of large pulses shows that nothing analogous to black hole formation occurs.  We have argued that the entire scattering matrix consists of the 't Hooft phase shift, convoluted with the transformations between the near horizon coordinates $u,v$ and the ingoing and outgoing fermion coordinates $t \pm r$.  

We will try to argue that the semi-classical analysis is valid only in the deformed model and the large $N$ limit.  Briefly, the argument is the following: Send in a pulse consisting of $o(N)$ fermions.  We can then occupy the meta-stable levels with a large number of distinct fermion flavors, creating a meta-stable system with energy levels
\begin{equation} E = - \sum n_i E_i  .  \end{equation} Since the $E_i$ are relatively irrational and the $n_i$ are large, the level structure is very chaotic. It will obey Poisson statistics for large $N$.

The time scale for single fermion emission is $\omega^{-1}$, independent of $N$.  If the emission is thermal, then the temperature is $N$ independent, while the entropy is $o(N)$, and the lifetime of the metastable state is $o(N)$.
These are, of course, the characteristics of a linear dilaton black hole.
The only question that remains is whether the emission of fermions from these meta-stable states is indeed thermal, because the system has an infinite number of conservation laws.  It is possible that highly entangled states of the system, in which none of the conservation laws but energy are diagonal, still exhibit thermal behavior, but to remove this issue we modify the model further by taking a relevant perturbation of the form
 \begin{equation} \delta H = \int d\lambda \psi^{i\ \dagger} (\lambda ) V_{ij} (\lambda ) \psi^j (\lambda) . \end{equation}  The matrix valued potential $V_{ij} (\lambda )$ satisfies
 $ [V(\lambda ) , V(\lambda^{\prime}) ] \neq 0$ and the commutator has full $N \times N$ rank.  It seems extremely plausible that the emission of fermions by the meta-stable states is thermal for this model, so that they have all the characteristics of linear dilaton black holes.
 
 Note that there is no explicit trace of the meta-stable states in the low energy effective field theory.  They are all at Planck energies, though their splittings are much smaller than this for large $N$.  To understand the spatial distribution of the states, we revert to the model with $V_{ij} = \delta_{ij} V$, and compute the time of flight to some point $\lambda$ in the ground state
 \begin{equation} \omega t(\lambda) = \int_0^{\lambda}\ \frac{dy}{\sqrt{- V(y) + y^2}} , \end{equation} and the entropy of that region.  
 \begin{equation} S (\lambda) = N \int_0^{\lambda}\ dy \sqrt{- V(y) + y^2} . \end{equation}
 
 We've used the Thomas-Fermi/WKB approximation to compute these quantities.  The entropy in this calculation is roughly the logarithm of the number of available states near the ground state, which may be considered localized in the interval $|y| \leq \lambda$.   A more precise definition, in terms of the entanglement entropy of the region $[ - \lambda, \lambda ]$ in the ground state, has recently been given in \cite{das}\cite{hartnoll} . Since the time of our Hamiltonian system is the same as that in the Minkowski space-time of our model, and the fermions are massless in that space-time, the time of flight tells us the value of the spatial coordinate corresponding to $\lambda$.  This leads to the relation $2 S(r) = N e^{2 r M_P} $, which is the $\mu$ goes to zero limit of the familiar matrix model relation\footnote{The transformation between $\lambda$ and $r$ is said to be non-local in the string theory literature.  This appears to be a stringy effect, which goes away when $\mu$ is small.  It may be a relative of stringy corrections to the black hole entropy formula in higher dimensions.}.  Changes in the potential near $\lambda = 0$ lead to $o(N)$, $r$ independent, changes in entropy, which can be compensated by an $o(1)$ shift of $r M_P$.  This is consistent with the fact that they shift the time of flight by a few Planck units.  We conclude that any meta-stable states that can be created by such a change of potential are localized within a few Planck lengths. 
 
 The time of flight calculation also tells us something about the interior of the black hole.  Incoming fermions encounter an environment different from the linear dilaton vacuum for only a few Planck times after they ``cross the horizon".  Horizon crossing is a somewhat emergent and nebulous concept in our models.   As we reviewed above, if we send in a pulse of $o(N) $ fermions, the classical gravitational equations develop a horizon at a position where the entropy function of the linear dilaton vacuum is $o(N)$.  This might suggest a length scale of order ${\rm ln}\ N L_P$ in black hole physics. However, as we've noted, the classical in-fall time to the singularity is $o(1)$, independent of $N$.  
 
 The exact quantum mechanics of the model shows no long time scale either.   If we look at processes in which just a few left moving fermions are excited, so that no black holes are formed, we see large phase shifts when the incoming momenta are close to the Breit-Wigner resonances of the single body S-matrix.  The time delays are $o(1)$ in Planck units.  
 This is unchanged by exciting a black hole.  The details of the scattering depend on the particular model we choose.  The model with $V_{ij} \propto \delta_{ij} $ conserves individual fermion numbers.  So if we throw in a particular fermion, of type $i$, with a wave packet which reaches the dip in the potential long after the black hole is formed, then it will see a large phase shift, or not, depending on the probability that this particular fermion's meta-stable state is ``still occupied" at that point in the history of black hole decay.  Note that a generic micro-state of the black hole, with a lot of entanglement between different types of fermion, will, at a time the black hole has large entropy, have probability $\sim 1/2$ for this particular fermion's meta-stable level to be occupied.
 
 The model with a matrix valued potential $V_{ij}$  will have behavior that is more like the behavior we expect for higher dimensional black holes.  Individual fermion numbers are not conserved and we no longer have to avoid special initial states which will exhibit non-thermal behavior.
However, although these models appear to be completely sensible perturbations of a well established model of gravitational scattering theory, they may appear artificial in that they exploit the lack of translation invariance of the $1 + 1$ dimensional linear dilaton vacuum.  The modifications which determine whether or not a model has black holes have to do with the Planck scale region where the string coupling goes to infinity\footnote{It should be clear to all readers that the behavior of these models with respect to the existence and properties of black holes will persist when we turn on the string coupling/shift the fermi level away from its CP symmetric value.  For weak string coupling, black hole production and decay will be a non-perturbative effect.  }.

\subsection{Two Body Deformations}

We can make models which partially remove this defect by replacing the deformation of the potential with a localized four fermion interaction

\begin{equation} \delta L = \int d\lambda d\kappa \psi_i (\lambda ) \psi_i (\lambda ) M_{ij} (\lambda , \kappa) \psi_j (\kappa ) \psi_j (\kappa ) . \end{equation}
$M_{ij}$ is a quasi-local kernel, localized near the top of the upside down oscillator potential.  An example would be a linear combination of the $r (\beta_{ij}, k_{ij} )$ .  Note that there is no reason to insist on a symmetry between the different fermion flavors.  If we imagine these models as arising from higher dimensional linear dilaton black holes in a variety of compactifications of string theory, with the massless fields coming from the Ramond-Ramond sector, then complicated compactifications can lead to many such fields, associated with different singularities of the internal manifold.  No continuous symmetries relate them.

If the number of fields is large, then the Hartree approximation will give a good estimate of the properties of the system.  If in addition the density density interactions are attractive\footnote{One can guess that repulsive interactions might lead to phase shifts exhibiting {\it time advances}, rather than {\it time delays}, violating causality from the point of view of the relativistic fermions $\Psi_{in/out}$.  Note that there would be nothing wrong with time advances in the non-relativistic model.  This question deserves further study.} then the Hartree equations will have Lee-Wick\cite{leewick} soliton solutions describing meta-stable states of fermions in a self-consistently generated potential.

Such models would avoid all issues of integrability and are more satisfying  because the black holes they contain are generated by interactions between the particles that form them, rather than a {\it potential ex machina} .   The absence of Lorentz invariance in all $1 + 1$ dimensional gravitational scattering theories, makes it harder to formulate axioms delineating which models are or are not acceptable.  Apart from causality constraints (see previous footnote), and the preservation of the correct asymptotic behavior,
we have only the general constraints of time translation invariance and unitarity to guide us.  All the models we have proposed obey these constraints.

\subsection{Global Conservation Laws}

Part of the folklore of quantum gravity is that there are no global continuous symmetries.  There are two very different kinds of arguments for this proposition.  The first class of arguments comes from perturbative string theory in Minkowski space and the AdS/CFT correspondence. One uses Noether's theorem in the world sheet Lagrangian or the boundary CFT to construct a world sheet/boundary current.  The world sheet current is used to create a vertex operator for a gauge boson, and the boundary current is dual to a bulk gauge boson.   

The second class of arguments comes from black hole physics. If the minimal mass of a globally charged particle is $M < M_P$ one can form a black hole of mass $NM$ and global charge $N$, for arbitrarily large $N$.  These black holes will evaporate down to a mass of order $M$, by the Hawking process, which emits, on average, equal numbers of positive and negative charges.  Thus, the theory will have an infinite number of remnant particles, which fit inside a small Schwarzschild radius.   In dimension higher than $1 + 1$, this violates the covariant entropy bound.  We can see two immediate reasons why this argument does not work in our models.  Most importantly, the connection between entropy and geometry is broken.  The entropy function is just the dilaton field, which has no relation to geometrical data.  In addition, in our models, the global conservation law is a charge counting massless fermions. 

The actual process by which our black holes decay is emission of massless particles.  If the entropy is much larger than the charge, this emission will begin in a roughly thermal manner with the average charge emitted small compared to the entropy change.  However, even if the system reaches the ``extremal limit" in which all the meta-stable states are filled with fermions rather than anti-fermions it will continue to decay by emitting massless fermions of a single sign of charge.  Note by the way that in the model where black holes are formed by an attractive density density interaction, the condensate found in the Hartree approximation carries a net charge, of order its entropy.  It will decay by fermion emission.

The loophole in the world sheet/ads/cft arguments sheds some light on the above description.  Those arguments alone, do not specify the Lagrangian of the bulk gauge field, which has to be constructed in order to match S-matrix/boundary correlator calculations.   In $1 + 1$ dimensions, a possible Lagrangian for a bulk $U(1)$ gauge field is the topological term
\begin{equation} L = \chi F , \end{equation} where $F$ is the gauge field strength and $\chi$ a scalar.  When the gauge potential is coupled to the fermion number current $J^{\mu}$ we get the equation of motion
\begin{equation} J^{\mu} = \epsilon^{\mu\nu} \partial_{\nu} \chi , \end{equation} which just says that 
$\chi$ is the bosonized fermion field.  The total charge is related to the value of $\chi$ at infinity.  

Just as the dilaton has to be considered a ``part of the gravitational field", we should think of $\chi$ as ``part of the Maxwell field".  Like the dilaton, it is a constrained field, completely determined on shell in terms of the dynamical matter fields.  From this point of view we see that the mantra that ``every continuous symmetry is gauged" is valid in our $1 + 1$ dimensional models as well.  The fermion number current is the curl of a gauge potential, whose value at infinity determines the total charge.  The Lagrangian $\int \chi F$ has a gauge symmetry under constant shifts of $\chi$ when $\int F = 0$, which becomes a discrete shift in quantum mechanics if $\int F$ has quantized values.  Note finally that the Hartree condensate of our Lee-Wick black holes in the 4 fermion model is, from this point of view, the Maxwell field of a charged linear dilaton black hole.

\section{Conclusions}

We have learned a number of general lessons from this excursion into $1 + 1$ dimensions.  Perhaps the most important is that 't Hooft's commutator algebra for near horizon null plane coordinates might be an exact feature of black hole scattering matrices.  As emphasized recently by Polchinski\cite{joechaos}, the actual derivation of these relations has very restricted validity within the realm of effective field theory.  There is a plausible derivation of the change in the black hole S-matrix when one adds a small number of particles to the scattering from a large black hole.  The leap from this to a second quantized theory of fields depending on the non-commuting null plane coordinates is huge, and not justified by deductive logic.

However, we've seen that the 0B $\hat{c} = 1$ model, which, along with its 0A cousins and BFSS Matrix Theory, is among our only examples of a non-perturbatively defined gravitational scattering theory, has an S-matrix that exactly reflects the 't Hooft commutator algebra.  This suggests that the algebra may be more intrinsic to quantum gravitational scattering than its derivation suggests.

One of the consequences of the algebra is that the classical notion of horizon is not sharp in the quantum theory.  The classical horizon is at $uv = 0 $.  The non-vanishing commutator implies that the closest quantum equivalent is $uv + vu = 0$.  In the $\hat{c} = 1$ model, this is obeyed only for one single particle eigenstate.  The ground state of the second quantized model has all of the single particle states with $uv + vu$ negative, filled.  Excitations of the fermi surface corresponding to localized incoming waves in the $r$ coordinate are states composed of fermions in superpositions of eigenstates of $uv + vu$. In the deformed models, which have black hole like excitations when $N$ is large, the meta-stable black hole micro-states are localized near $\lambda = 0$ .  $uv + vu$ is not sharp in these states, but the fluctuations of dimensionally correct light front coordinates $(U,V) = L_P (u,v)$ are of order Planck scale.  Note that large $N$ condensates of fermion bilinears {\it will} depend on sharply defined $u,v$ coordinates.   This is analogous to the sharpness of soliton position and velocity operators in weakly coupled quantum field theories.

We've emphasized that none of these models enables us to address the AMPS paradox, because the classical predictions for linear dilaton black holes of arbitrarily large entropy do not imply a parametric difference between the in-fall time to the singularity, and the Planck time.  Planck scale uncertainties in the position of the horizon, also do not provide evidence falsifying the effective field theory treatment of horizon dynamics.  The only definitive evidence against that treatment in the models we have proposed is that  effective field theory is insensitive to the existence of Planck scale deformations of the fermion potential, which differentiate between models with and without black holes.   A picture of a substantial portion of black hole entropy as coming from effective field theory states on a stretched horizon in Schwarzschild coordinates is at odds with the existence of different models with the same effective field theory, only some of which have actual black hole excitations.

We've noted that the entropy independent in-fall time for linear dilaton black holes is consistent with the assertion in \cite{fw23} that dynamics inside the horizon of a higher dimensional black hole is the mirror or mirage of the dynamics of scrambling on the horizon, as seen by an external observer.
A {\it highly speculative} model, which connects the work of \cite{fw2,fw3} to the present paper, is given by a Hamiltonian of the form
\begin{equation} H_{spec} = \int du (\psi^{\dagger\ i}_A (u) (uv + vu) \psi_i^A (u^{\prime})+ \frac{1}{N} {\rm Tr} P(M ) , \end{equation} where $M_i^j (u,u^{\prime}) = \frac{1}{N} \psi_i^A (u) \psi^{\dagger\ j}_A , $ and the ``polynomial" $P$ involves $u$ dependent coefficients, which go to zero at large $u$.  The Tr operation involves integrals over $u$. The $\psi$ is, for each value of $u$, an $N \times N+1$ matrix, where $N$ is proportional to the Schwarzschild radius of a four dimensional black hole.  There's a corresponding representation of the same Hamiltonian in terms of the $v$ coordinate, where the $\psi_i^A (v)$ fields are related to $\psi_i^A (u)$ by Fourier transform.

The fantasy that goes along with $H_{spec}$ is that this Hamiltonian represents the near horizon S-matrix, $S_{hor}$ in the language of 't Hooft\cite{tH}, for a four dimensional Schwarzschild black hole. $u,v$ are the 't Hooft near horizon light front coordinates.  We should note that 't Hooft argued for the existence of light front coordinates $u (\Omega ) , v (\Omega )$ localized on the spherical holographic screen of the black hole horizon, with commutator 
\begin{equation} [u (\Omega ) , v (\Omega^{\prime} ) ] = \frac{1}{L^2} (\Omega , \Omega^{\prime} ) . \end{equation} We've treated the sphere by cutting off the angular momentum operator $L^2$ acting on its spinor bundle.  The angular momentum decomposition of the commutator above just gives Kronecker deltas, multiplied by $\frac{1}{l(l + 1)}$ and we've renormalized them so that they all have the same normalization, and pushed the angular momentum labels to the second quantized fields.  Note that although the system described above has infinite entropy, that entropy should be associated with scattering states, rather than the black hole itself.  This is precisely analogous to what happens in the $1 + 1$ dimensional models. The black hole entropy is associated with meta-stable states localized in the near horizon regime where $u$ and $v$ are small.

The matrix Hamiltonian is assumed \cite{sekinosusskind} to have fast scrambling behavior, which models the decay of localized perturbations of the black hole horizon.  The natural time scale of this decay is $o(N)$ and the scrambling time is $o(N {\rm ln}\ N)$ in Planck units.  The Fourier transform relation between the $u$ and $v$ descriptions shows that both of them will exhibit the same behavior.  We can interpret this either as scrambling on the horizon as seen in an outgoing coordinate system, or as the experience inside the horizon as seen in an in-falling coordinate system.  This is consistent with the claims of \cite{fw23}.  It need hardly be emphasized that, at this point, this model is just a speculative fantasy.

Returning to more concrete issues, the most pressing technical question raised by our work is to determine what the rules are, which govern allowed deformations of the $\hat{c} = 1$ model.  We believe in this model because of the existence of the marginal deformation of shifting the fermi level, which allows us to match it with a weakly coupled string expansion in the limit where the fermi energy is much larger than $\omega$.  On the other hand, the arguments of \cite{cghs} suggest that we should view it as a special case of a decoupled near horizon limit of linear dilaton black holes in higher dimensional string compactifications.  There are many such compactifications and many different extremal black holes in each compactification.  This can readily account for the existence of models with a wide variety of massless $1 + 1$ dimensional scattering states, which can always be mapped to a collection of left and right moving Weyl fermions.  The Alexandrov-Kazakov-Kostov map takes this into a collection of non-relativistic fermions in an upside down oscillator potential, which has a built in 't Hooft near horizon scattering matrix.
It seems unlikely that this is a unique and universal scattering matrix for all of these systems, especially since it does not have behavior corresponding to the black holes encountered in low energy effective field theory.  From the higher dimensional point of view, the non-existence of near extremal black holes seems particularly bizarre.  We've suggested deformations of the model which appear to contain meta-stable quantum states with the requisite properties.  All of these models preserve the asymptotic space-time picture and their low energy effective Lagrangians differ only in the number of Weyl fermions.  They are all consistent with the general rules of quantum mechanics, and obey the causal restriction that scattering exhibits only time {\it delays}.  They all have as much symmetry as one can expect of a linear dilaton system.  Are they all consistent models of quantum gravity?  Do they all arise as decoupling limits of black hole throats in higher dimensional string theory?  These questions remain for the future.  The 0A matrix models discussed in \cite{bose0AB} obviously have a structure similar to that discussed here (see the appendix to \cite{maldaseibflux} for a particularly lucid discussion), it would be worthwhile to work out the details and to understand relevant perturbations of those models.

Although we began this paper with a demonstration that the basic structure of the 0B matrix model could have been derived from the principles of HST, we have not elaborated the actual structure of an HST treatment of this model.
There are several reasons for this. HST was designed to deal with models of space-times whose class of distinguished time-like trajectories was infinite, and indeed to provide a framework in which one could in principle discuss local physics as seen from {\it any} time-like trajectory.  It was always clear that, for example, the physics of scattering in Minkowski space would be much simpler along Minkowski geodesics than along other trajectories.  In the linear dilaton vacuum, there is a preferred single parameter family of time-like trajectories, distinguished only by the choice of $ t = 0$ in the frame where $\partial_t \phi = 0$.  Indeed these are just different parametrizations of the same minimal entropy trajectory.  Correspondingly, scattering in these models occurs only near the minimal entropy point, so the minimal entropy trajectory is ``where the action is".  One would only want to introduce other trajectories in order to demonstrate that the model was generally covariant.  We've instead achieved that aim by relating the model to well established results in string theory\footnote{It's not inconceivable that the HST consistency relations for different trajectories could help to delineate the class of allowed relevant perturbations.}, and to 't Hooft's conjectures about the near horizon S-matrix.

Thus, the only thing that's really missing for a full HST description of these models is a nested tensor factorization of the Hilbert space, relating to larger and larger intervals in $t$.  This should treat the ingoing and outgoing coordinates symmetrically.  One interesting possibility is to expand the field operator, in Schrodinger picture in terms of annihilation operators for the 
single particle eigenstates of the harmonic oscillator $u^2 + v^2$ .  In terms of the creation and annihilation operators of that oscillator, our one body boundary Hamiltonian is proportional to $(a^2 + a^{\dagger\ 2})$.  The second quantized Hamiltonian is 
\begin{equation} H = \sum_n \psi_n^{\dagger} [c(n) \psi_{n + 2} + d(n) \psi_{n - 2} ] , \end{equation} where terms involving operators with negative indices vanish .  $c(n) = \sqrt{(n+1)(n+2)}$ and $d(n) = \sqrt{n(n-1)}$.  This looks like a hopping Hamiltonian on a one dimensional lattice, so it's easy to break it into $H_{in} (t) + H_{out} (t)$ , by omitting a single link.   Choosing the broken link to be at 
$n(t) = [e^{t\omega}] $ we get an entropy of the ${\it in}$ Hilbert space exponential in $t$, which is what we expect for a causal diamond of proper time $t$ in the linear dilaton vacuum.  

The $n-th$ eigenfunction of the $u^2 + v^2$ oscillator has its maximum in $\lambda$ at $\lambda_n^2 \sim n$ .   The time of flight map assigns this to a time $t_n = \frac{1}{2} {\rm ln}\ n$, which corresponds to entropy $n$ in the semic-classical approximation. 
The Hamiltonians $H_{in} (t)$ describe physics inside the causal diamond of the minimal entropy trajectory between times $[- t, t]$.  They correspond to a hyperbolic time slicing interleaved between the boundaries of causal diamonds.  In this model, choosing $H_{out} (t)$ to be ``the rest" of the boundary Hamiltonian with the link between $n(t)$ and $n(t) + 1$ broken, guarantees that in the limit $t \rightarrow \infty$ evolution with $H(t)$ approaches that of the time independent boundary Hamiltonian.  

In this simple model, we already had an effectively local description of $1 + 1$ dimensional space-time via the maps between asymptotic and eigenvalue coordinates.  The full machinery of HST is overkill, but it is comforting to see that it does exist.

There are a number of interesting questions raised by this work, to which we hope to return in the future.  Most importantly, one can hope that the lessons of this model will enable us to find an HST approach to models in AdS space.

 \vskip.3in
\begin{center}
{\bf Acknowledgments }
\end{center}
 The work of T.B. was supported in part by the Department of Energy. This paper was mostly conceived and entirely written, while the author was resident at KITP in Santa Barbara.  I would like to thank the Institute members and staff for their hospitality.  I'd also like to thank Willy Fischler for discussions related to this work.

\end{document}